\documentclass[journal,onecolumn,11pt]{IEEEtran}

\input packages-headers

\RequirePackage[top=3.0cm,bottom=3.0cm,left=2.5cm,right=2.5cm,nohead]{geometry}

\newcommand{\MATLAB}{MATLAB$^\text{\textregistered}$}
\newcommand{\Spectre}{Spectre$^\text{\textregistered}$}
\newcommand{\HSPICE}{HSPICE}

\definecolor{MyDarkGreen}{rgb}{0.0,0.4,0.0} % This is the color used for comments
\definecolor{MyDarkRed}{rgb}{0.4,0.0,0.0} % This is the color used for strings
\renewcommand{\matlabscript}[3]{
    \begin{itemize}
    \item[]\lstinputlisting[language=Matlab,caption=#3,label=#2]{#1}
    \end{itemize}
}

\newcommand{\verilogascript}[3]{
    \begin{itemize}
    \item[]\lstinputlisting[language=C,caption=#3,label=#2]{#1}
    \end{itemize}
}

\begin{document}
\renewcommand{\refname}{References}

\let\redHL=\ignore
\let\jnote=\ignore
\let\tnote=\ignore
\let\tsumm=\ignore

\pagestyle{plain}
\thispagestyle{empty}

\title{Sub-harmonic Injection Locking in Metronomes}

\author
{
Tianshi Wang\\
{\small
Department of Electrical Engineering and Computer Sciences, University of
California, Berkeley, CA, USA}\\
Email: \texttt{tianshi@berkeley.edu}
\vspace{-1em}
}

\maketitle

\begin{abstract}
In this paper, we demonstrate sub-harmonic injection locking (SHIL) in
mechanical metronomes.
To do so, we first formulate metronome's physical compact model, focusing on
its nonlinear terms for friction and the escapement mechanism.
Then we analyze metronomes using phase-macromodel-based techniques and show
that the phase of their oscillation is in fact very immune to periodic
perturbation at twice its natural frequency, making SHIL difficult.
Guided by the phase-macromodel-based analysis, we are able to modify the
escapement mechanism of metronomes such that SHIL can happen more easily.
Then we verify the occurrence of SHIL in experiments.
To our knowledge, this is the first demonstration of SHIL in metronomes;
As such, it provides many valuable insights into the modelling, simulation,
analysis and design of nonlinear oscillators.
The demonstration is also suitable to use for teaching the subject of injection
locking and SHIL.
\end{abstract}

\thispagestyle{empty}
\section{\normalfont {\Large Introduction}}\seclabel{intro}

Injection locking is an interesting phenomenon observed in almost all nonlinear
oscillators.
When an oscillator with natural frequency $f_0$ is perturbed by a small
external periodic input at a frequency $f_1$ ($f_1 \approx f_0$), the
oscillator will forget about its natural frequency and move its energy to lock
to $f_1$.
Because of injection locking, coupled oscillators often synchronize in both
frequency and phase.
Many natural phenomena can be explained from this mechanism, such as the
synchronization of fireflies' patterns, neurons firing in unison, \etc{}
% newcommands-eg-ie-etc.tex
Injection locking is also widely used in the design of radio frequency
communication circuits and optical lasers.

Perhaps the most famous example of injection locking in oscillators is the
demonstration of metronomes synchronizing their ticks.
As illustrated in \figref{metronome-sync}, when several metronomes are placed
on a platform that can roll horizontally, each of them receives a small
perturbation from their neighbours through the common platform.
They may have slightly different central frequencies and may be started with
random phases.
But given some time, through the mechanism of injection locking, all of them
will eventually lock to the same frequency with the same phase.
This synchronization phenomenon is reproducible, easy to see and hear.
As such, it is often used to illustrate or teach the subject of injection
locking.

\begin{figure}[htbp]
\centering{
    \epsfig{file=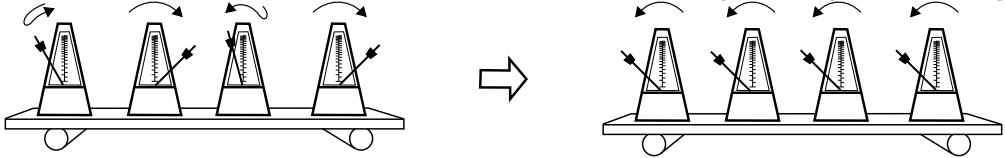,width=0.8\linewidth}
}
\caption{Metronomes standing on a rolling board end up ticking in unison.
    \figlabel{metronome-sync}}
\end{figure}

Sub-harmonic injection locking (SHIL) is a special type of injection locking.
Under SHIL, oscillators are perturbed by a faster periodic input with a $2f_1$
frequency ($f_1 \approx f_0$).
As illustrated in \figref{SHIL}, the oscillator will lock to $f_1$, which is
the sub-harmonic of the perturbation.
SHIL has wide-ranging applications, such as in the design of high-performance
quadrature oscillators \cite{kinget2002injection}, injection-locked PLLs
\cite{Abramovitch2002ACC}, frequency dividers \cite{tiebout2004cmos}, optical
lasers \cite{GoTaWeBl1983Lasers}, \etc{}
Recently, it has been shown that, through this mechanism, an oscillator can
develop bistable phase responses that are suitable for encoding and storing
phase-based logic bits for Boolean computation
\cite{WaRoUCNC2014PHLOGON,RoPHLOGONarXiVv12014}.
This new computation paradigm has potential noise and power advantages over
conventional level-encoded computation schemes.
In this context, SHIL has been demonstrated in CMOS ring oscillators and LC
oscillators
\cite{WaRoUCNC2014PHLOGON,WaRoDAC2015MAPPforPHLOGON}.

\begin{figure}[htbp]
\centering{
    \epsfig{file=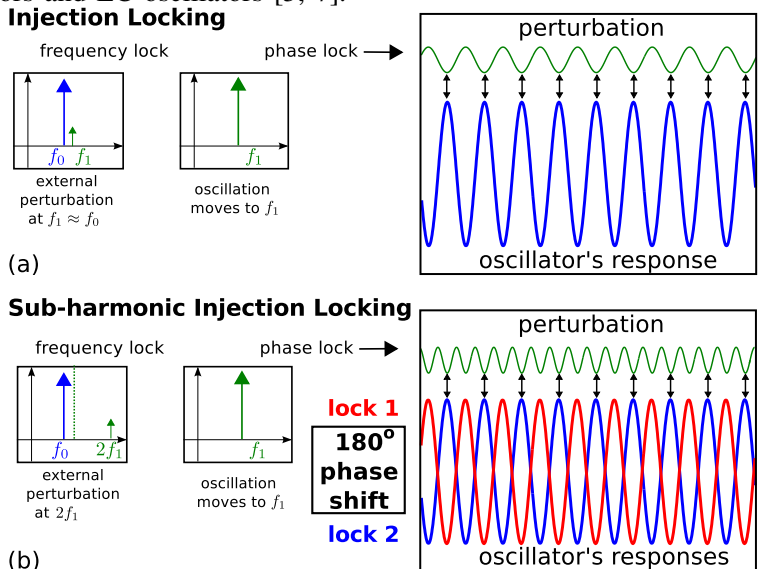,width=0.6\linewidth}
}
\caption{Sub-harmonic Injection Locking implies phase lock with
         multiple stable phases.
		 (a) When perturbing a self-sustaining oscillator (\eg, a
		 metronome) with input at $f_1$ close to its natural frequency $f_0$, 
		 the oscillator moves its energy to $f_1$ and features phase lock. 
		 (b) When the perturbation is at $2f_1$, the oscillator may move to
		 $f_1$ with bi-stable phase locking states.
    \figlabel{SHIL}}
\end{figure}

However, SHIL has never been demonstrated in mechanical metronomes.
The basic idea of such a demonstration is simple: just like in the scenario of
regular injection locking (we refer to it as IL in the rest of the paper), two
metronomes are placed on a rolling board --- one oscillates at approximately
the $1/2$ sub-harmonic of the other.
If SHIL happens, their swing patterns will synchronize.
To complete this demonstration, they should also be decoupled by stopping the
rolling board, in which case the synchronization ceases.
Such a demonstration can show that SHIL is not limited to electrical
oscillators --- it is almost as universal as regular IL.
It can also serve as an eye-catching illustration when teaching the subject of
SHIL in classes.

The lack of such a demonstration is not because of lack of trying.
In fact, we have been attempting to demonstrate it in our group for years.
But simply tuning two metronomes to 1Hz and 2Hz\footnote{Frequencies used in
this paper are the frequencies of metronomes' oscillation. One cycle of
metronome oscillation generates two ticks. So 1Hz and 2Hz metronomes generate
120 and 240 beats per minute respectively.}
and putting them on a rolling board does not result in synchronization.
The coupling seems to have little effect and the detuning in their frequencies
remains, separating their phases apart.
The metronomes we use\footnote{Wittner Taktell Super-Mini Metronome} oscillate
for approximately 17 min at 2Hz; clear and secure synchronization appears
impossible to achieve within this time frame.
This observation leaves us an impression that metronomes are very immune to
perturbation close to its natural oscillation's second harmonic (we refer to it
as second-order perturbation in the rest of this paper).
This intuition needs more concrete analysis and justification.

In this paper, we adopt a more rigorous approach.
We start with writing the physical model of metronomes in \secref{model}.
Then in \secref{PPV} we use phase-macromodel-based techniques to analyze the
model and debug why SHIL is hard to achieve.
Then guided by the phase-macromodel analysis, we tweak the metronome, adjust
its PPV such that SHIL can happen more easily.
Finally, in \secref{results}, we validate the occurrence of SHIL in metronomes
by experiments and measurements.
As described above, we put a 1Hz metronome and a 2Hz metronome on a rolling
board so that they can be coupled to each other; then we decouple them by
putting them on a stationary board instead.
In both scenarios, we tape colored labels to the rods of the metronomes, record
videos and process them.
\figref{photos-Lissajou} shows the experimental setup and the key results.
The curves in \figref{photos-Lissajou} (b) and (d) are known as the Lissajous
curves --- they plot the x coordinates of the red dot \wrt{} those of the blue
dot.
On a rolling board, Lissajous curves in \figref{photos-Lissajou} (d) clearly
show a pattern, indicating that the two metronomes synchronize.
In the case of the stationary board, because of detuning, the Lissajous
curves span the whole plane within the swings of oscillation.

\begin{figure}[htbp]
\centering{
    \epsfig{file=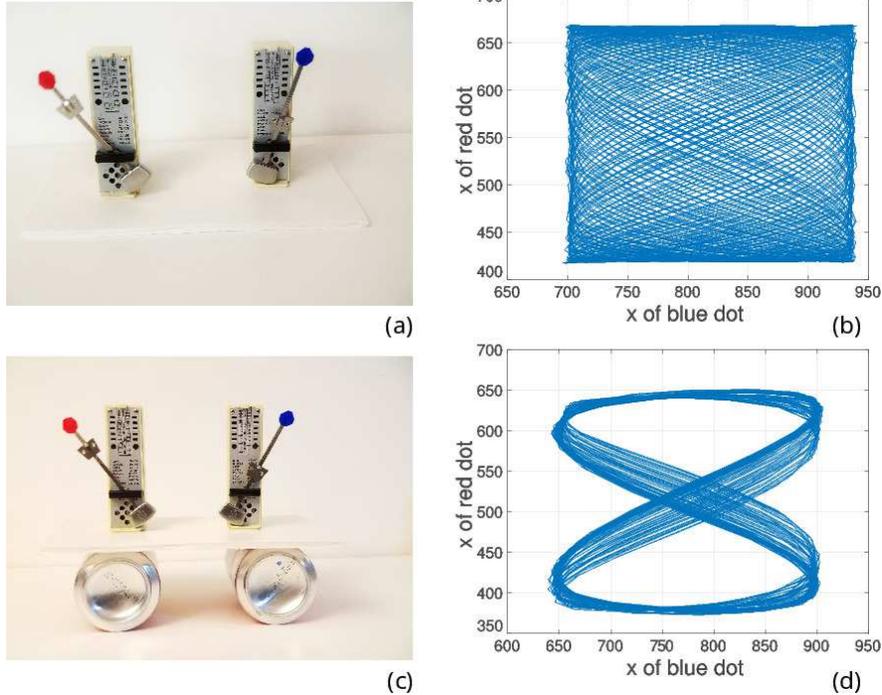,width=0.7\linewidth}
}
\caption{(a) Two metronomes tuned to approximately 1Hz and 2Hz are placed on a
stationery platform; (b) the corresponding Lissajou curves in show that no
synchronization happens; (c) the metronomes are coupled through a rolling
board; (d) the pattern in Lissajou curves demonstrates SHIL.
    \figlabel{photos-Lissajou}}
\end{figure}

Along with the key results, this paper also presents compact models for
metronomes that can run in open-source and commercial simulators, together with
useful modelling techniques.
Metronomes are normally modelled as lossless double-weighted pendulums
\cite{pantaleone2002metronome,jia2015triplet}.
But the ``lossiness'' of an oscillator actually plays an important role in its
injection locking behaviour.
An accurate metronome model needs to include friction damping and also a
mechanism known as escapement that compensates the energy lost every cycle due
to friction.
The implementations of the several existing metronome models with these
mechanisms are not openly available.
Neither are they formulated to be compatible with main-stream simulators.
Moreover, friction and the escapement mechanism are often modelled in these
models using non-smooth functions, making them not suitable for simulation.
In our model, we use smooth functions to alleviate the difficulty of simulation
convergence.
As we explain in \secref{model}, the use of smooth functions actually increases
the models' physical fidelity as well.
The metronome models presented in this paper are implemented in both
ModSpec/MATLAB and Verilog-A, and open-source released at \cite{metronomewebsite}.
In the Verilog-A version, we use the regular electrical discipline and model
kinematic quantities as voltages, such that the model can be usable in as many
simulators as possible.
% In the Verilog-A version, we demonstrate both the use of kinematic discipline
% and the regular electrical one for the model to be accepted in as many
% available simulators as possible.

\ignore{
Although metronome models exist in literature, they are done at the system
equation level, not from a compact modelling (device modelling) perspective.
And they are certainly not done in industry  standard Verilog-A language. We
have implemented Verilog-A models, using both EE discipline for compatibility
and kinematic discipline for more natural representation. The framework we have
proposed allows for simulation of more than just two metronomes, in proper
simulators for circuits. Moreover, this is the first time
phase-macromodel-based analysis has been applied to metronomes. PPV of the
metronomes is calculated numerically and accurately on realistic models,
providing insights into the modification of the metronome for SHIL. Then,
guided by insight, we are able to informedly modify the metronome, and for the
first time, demonstrate SHIL in metronomes.
}

To our knowledge, this is also the first time metronomes are analyzed using
phase-macromodel-based techniques.
Specifically, we extract a time-varying periodic vector --- perturbation
projection vector (PPV), from metronome model equations.
PPV describes the sensitivity of the metronome's phase to external perturbation.
As we show in \secref{PPV}, the fundamental frequency component and second
harmonic of the PPV indicate how easy it is for IL and SHIL to happen.
From the PPV numerical results, we confirm that SHIL is unlikely to occur
without any modification to the metronome.
And guided by the PPV results, we can easily tweak the escapement mechanism
such that SHIL can reliably happen.

The techniques we present in this paper on the modelling and analysis of
metronomes, especially the use of phase-macromodels to predict, analyze and
achieve desirable IL properties, are applicable and useful for the design of
almost any oscillator.
Moreover, the demonstration of SHIL shows that SHIL is a general phenomenon in
nonlinear oscillators.
Through this mechanism, a wide range of oscillators can potentially be used as
phase-based binary latches --- not just the electrical ones, but also those in
optics, MEMS, bio-chemistry, spintronics, \etc{}
Last but not least, the simulation and experimental results we present in this
paper, are all easily reproducible.
Just as metronomes are often used in teaching the subject of IL, the
demonstration of SHIL in metronomes can also be used for teaching the subject
of SHIL as well as oscillator modelling and design. 

\section{\normalfont {\Large Modelling Metronomes}}\seclabel{model}

The dynamics of a metronome are mainly governed by the equation of a
double-weighted pendulum. The equation is written using the angle $\theta$ and
angular velocity $\dot \theta = \frac{d}{dt} \theta$ of the
pendulum\footnote{The angle is defined between the pendulum and the vertical
position. The direction does not change any equations.}:
\begin{equation}
\frac{d}{dt} \dot \theta =
- \frac{1}{m_1 h_1^2 + m_2 h_2^2}
\cdot \left(m_1 g\sin(\theta) \cdot h_1 - m_2 g\sin(\theta) \cdot h_2 \right),
\end{equation}
where $m_1$ is the mass of the weight at the bottom of the pendulum,  $h_1$ is
the distance from it to the axis of rotation;
$m_2$ and $h_2$ are the mass and distance for the weight on the top of the
pendulum; $g$ is the gravitational constant.

Next, we add several terms to this pendulum equation --- friction, the
escapement mechanism, and external force.

Frictional forces in this system come from several sources: air friction, the
axle bearing, and the contact between the tooth of the escapement wheel and the
circular plate of the actuating member attached to the axle.
The first one --- air friction is often modelled as a viscous friction that
grows linearly with velocity.
The latter two are in the form of surface friction, which is often considered
as a constant force in the opposite direction of the relative movement.
They are much larger than the air friction in a metronome.
Therefore, we write the formula for friction as
\begin{equation}\eqnlabel{ffunc}
f(\dot \theta) = - f_0 \cdot \text{smoothsign}(\dot \theta),
\end{equation}
where $f_0$ is a fitting parameter representing the constant amplitude of
friction.

A double-weighted pendulum with only frictional forces will have damped
oscillation.
To sustain the oscillation, a metronome has a spring box inside that drives an
escapement wheel through gears.
The wheel has sloped teeth. At almost any time, one of the teeth is pushing
against a circular plate attached to the axle of the pendulum.
As the pendulum swings to certain angles where the circular plate is designed
to have a gap, the wheel ``escapes'' through the plate and moves one tooth
forward.
In the meanwhile, the movement of the tooth pushes the circular plate, making
the pendulum swing faster.
At the same time, a sound of a tick is generated.
This escapement happens twice during a cycle of oscillation, one when the
pendulum is swing left and the other right.
The angles at which the escapement occurs at left and right are normally
symmetric.
The pushing force on the pendulum generated by escapement is modelled as a
$g(\theta,~\dot \theta)$ function in our model, which is non-zero only at small
windows of $\theta$, with direction aligned with the sign of $\dot \theta$.
\begin{equation}\eqnlabel{gfunc}
\begin{split}
g(\theta,~\dot \theta) = &+g_0 \cdot (\text{smoothstep}(\theta-\theta_R) -
\text{smoothstep}(\theta-\theta_R-\Delta\theta)) \cdot \text{smoothstep}(\dot
\theta) \\
&- g_0 \cdot \left(\text{smoothstep}(\theta-\theta_L+\Delta \theta) -
\text{smoothstep}(\theta-\theta_L)\right) \cdot \text{smoothstep}(- \dot
\theta).
\end{split}
\end{equation}

The formula in \eqnref{gfunc} sets $g(\theta,~\dot \theta)$ to be about $g_0$
when $\theta_R < \theta < \theta_R +
\Delta\theta$ and $\dot \theta > 0$; $g(\theta,~\dot \theta)$ is about $-g_0$
when $\theta_L - \Delta\theta < \theta < \theta_L$ and $\dot \theta < 0$.

Furthermore, the external force applied to a metronome can be written as a
horizontal acceleration $a$ of the axle of the pendulum.

Putting together all the components discussed above, we have a metronome model
as follows.
\begin{equation}\eqnlabel{metronome}
\begin{split}
\frac{d}{dt} \dot \theta =
& - \frac{1}{m_1 h_1^2 + m_2 h_2^2}
\cdot \left(m_1 g\sin(\theta) \cdot h_1 - m_2 g\sin(\theta) \cdot h_2 \right.\\
& \left. + f(\dot \theta)
+ g(\theta,~\dot \theta)
+ m_1\cdot a\cdot \cos(\theta)\cdot h_1 - m_2\cdot a\cdot \cos(\theta)\cdot
h_2 \right).
\end{split}
\end{equation}

Among the parameters used in this model, $m_1$, $m_2$, $h_1$ and $h_2$ can be
directly measured using a scale and a ruler;
values for $\theta_R$, $\theta_L$ and $\Delta \theta$ can be obtained by
measuring the gap in the circular plate using a protractor.
$f_0$ is the nominal value of friction.
We can estimate it by letting the pendulum swing within small angles, such
that the escapement does not happen.
In this case, the metronome begins damped oscillation due to friction.
We can tweak the value of parameter $f_0$ until the simulated response matches
observation in the speed of damping.
$g_0$ is the force the escapement wheel applies to the pendulum during each
tick.
When all the other parameters are fixed, $g_0$ determines the swing of the
metronome.
We can estimate this parameter by matching the magnitude of oscillation in
simulation with the actual swing of the metronome.
From above, we have been able to systematically determine all the parameters in
the metronome model \eqnref{metronome}, either directly or indirectly.

Note that in the model equations for friction \eqnref{ffunc} and escapement
mechanism \eqnref{gfunc}, we use the smooth versions of sign and step
functions \cite{wang2016well}.
This is not just for improving the convergence of numerical simulation.
The use of smooth functions also makes the model represent the physical system
more truthfully.
For example, when the tooth of the escapement wheel starts to push the circular
plate, the force is not instantaneous because the surface at the edge of the
gap is still smooth.
Similarly, representing friction using smooth functions also increases the
model's physical fidelity \cite{makkar2005friction}.

\subsection{Metronome Model in MAPP}\seclabel{MAPP}

To run simulation algorithms on this metronome model, we implement it in MAPP,
the Berkeley Model and Algorithm Prototyping Platform.
MAPP models general continuous-time dynamical systems as differential algebraic
equations (DAEs).
A DAE has the following format:
\begin{equation}
\frac{d}{dt} \vec q(\vec x(t), \vec u(t)) + \vec f(\vec x(t), \vec u(t)) = \vec 0.
\end{equation}

In the metronome model, $\vec x = [\theta, \dot \theta]^T$, $\vec u = [a]$.
By specifying functions $\vec q$ and $\vec f$, we can write this model in the
DAE format.
\begin{eqnarray}
\vec q(\vec x(t), \vec u(t)) &=& 
-\left[ \begin{array}{c}
\theta \\
\dot \theta
\end{array}\right] = -\vec x\\
\vec f(\vec x(t), \vec u(t)) &=& 
\left[ \begin{array}{c}
\dot \theta \\
\frac{d}{dt} \dot \theta \text{ in equation \eqnref{metronome}}
\end{array}\right]
\end{eqnarray}

MAPP provides a convenient way of coding this model in the \MATLAB language.
The code is shown in \appref{metronome_DAE_code}.

\begin{figure}[htbp]
\centering{
	\epsfig{file=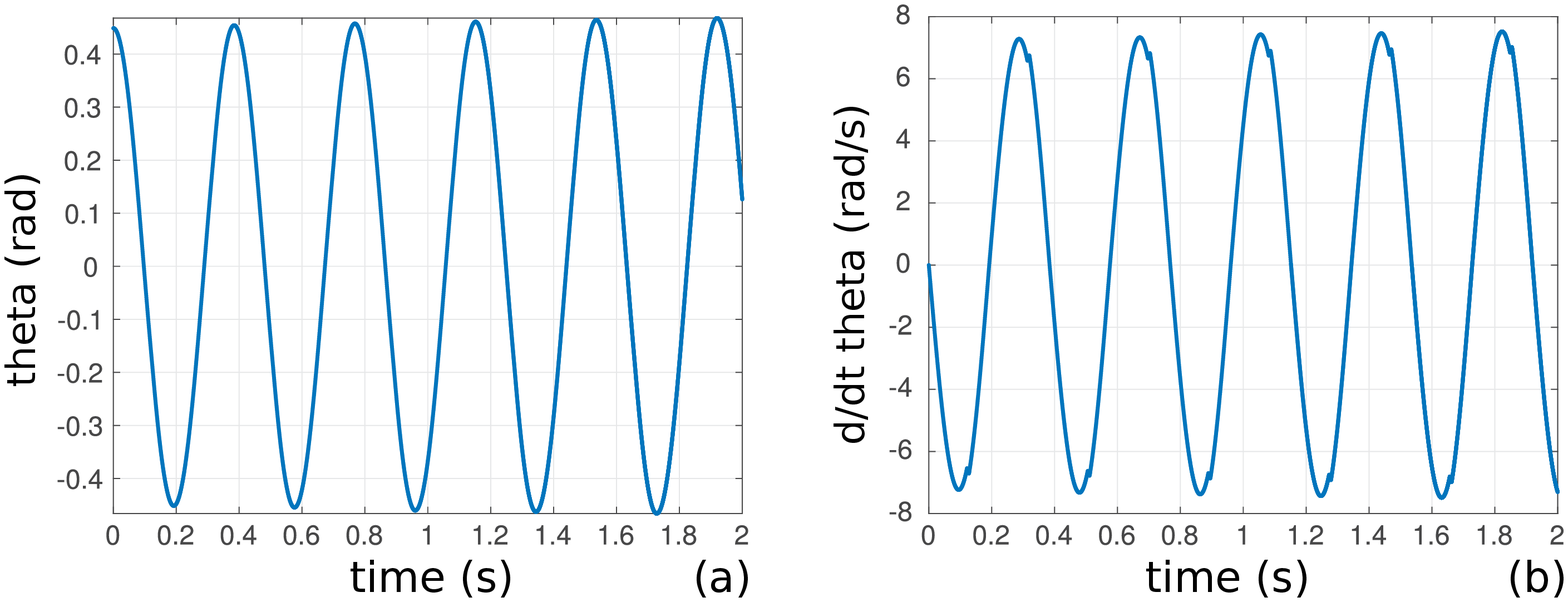,width=1.0\linewidth}
}
\caption{Transient simulation results of $\theta$ and $\dot \theta$ from MAPP.
\figlabel{metronome-waveforms}}
\end{figure}

\begin{figure}[htbp]
\centering{
	\epsfig{file=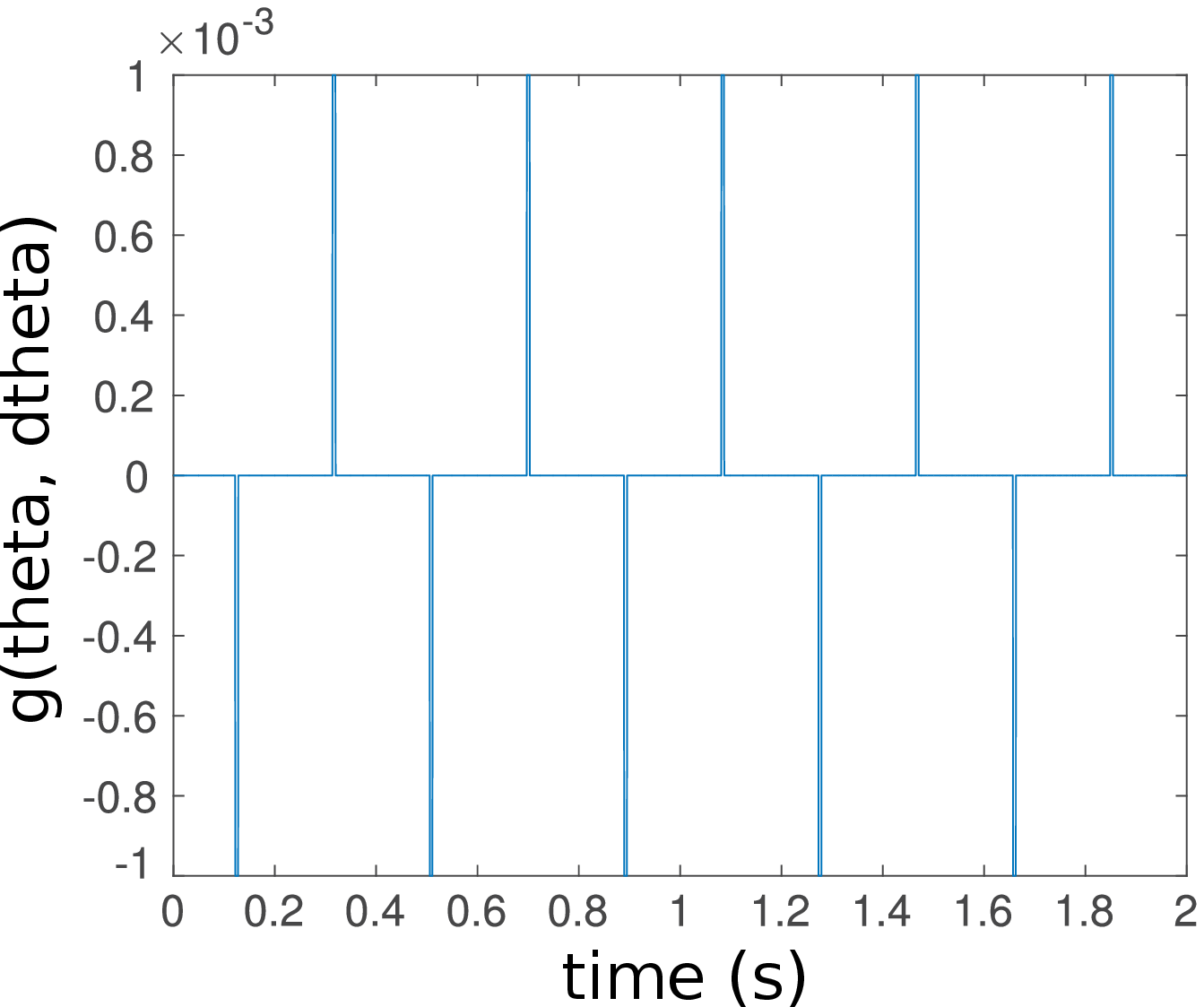,width=0.5\linewidth}
}
\caption{Transient simulation results of $g(\theta,~\dot \theta)$ from MAPP.
\figlabel{metronome-gfunc-waveform}}
\end{figure}

Results from transient analysis in \figref{metronome-waveforms} show that the
model reproduces the self-sustaining oscillation observed in metronomes.
Note that the waveforms of $\theta$ and $\dot \theta$ are not perfectly
sinusoidal, $\dot \theta$ actually has two small notches every cycle --- they
are where the escapement wheel moves forward one tooth.
Plot of $g(\theta,~\dot \theta)$ function in \figref{metronome-gfunc-waveform}
demonstrates this mechanism more clearly; unlike a spring-mass system or a
single pendulum, a metronome is indeed a nonlinear oscillator with highly
non-smooth dynamics.

\subsection{Metronome Model in Verilog-A}\seclabel{VA}

We can also implement the metronome model in the Verilog-A language and
simulate it in commercial circuit simulators.
Verilog-A is an industry-standard modelling language designed mainly for
electronic devices.
For the metronome model to run in most simulators, we do not use Verilog-A's
multiphysics disciplines;
we use only electrical domain constructs.
$\theta$ and $\dot \theta$ can be modelled as two voltages.\footnote{$\dot
\theta$ can be a current charging the $\theta$ node. But for printing and
plotting, it is more convenient to define it as a voltage.}
The Verilog-A code the for metronome model is shown in
\appref{metronome_va_code}.
It generates consistent results as in \figref{metronome-waveforms} and
\figref{metronome-gfunc-waveform} in commercial simulator \Spectre.

\ignore{
Available models for metronomes simulate with Matlab’s ODE solvers. They can
generate similar transient results. But our models are formulated inside
circuit simulators. As such, they can leverage many advanced simulation
algorithms developed mainly for circuits. Specifically, many oscillator
analyses for RF design can be applied, in both MAPP and commercial simulators
like Spectre. They wouldn’t have been done without proper modelling techniques.
}

\section{\normalfont {\Large Phase-macromodels of Metronomes}}\seclabel{PPV}

% Transient simulation is often not suitable for analyzing injection locking \cite{WaRoDAC2015MAPPforPHLOGON}.
In \figref{metronome-waveforms}, one can see that the waveforms' amplitudes
increase slightly with time; the oscillation has not settled to the limit
cycle yet.
Instead of waiting for it to settle in transient simulation, we can use other
algorithms to directly capture the limit cycle, \aka, the Periodic Steady State
(PSS).
There are two common PSS algorithms --- the shooting method, which is based on
transient analysis, and harmonic balance, which is a mixed frequency/time
domain method.
Both are implemented in MAPP and in \Spectre.
PSS analysis computes an oscillator's response $\vec x_s(t)$ such that
\begin{equation}
\vec x_s(t) = \vec x_s(t+T_0),
\end{equation}
where $T_0 = 1/f_0$ is the period of oscillation.

When the oscillator is under a small perturbation $\vec u(t)$, its oscillation
along the limit cycle develops a phase deviation $\alpha(t)$.
The oscillator's response under perturbation can be written as
\begin{equation}
\vec x(t) = \vec x_s(t+\alpha(t)),
\end{equation}
where $\alpha(t)$ is governed by a differential equation:
\begin{equation}\eqnlabel{PPV}
\frac{d}{dt} \alpha(t) = \vec v^T(t+\alpha(t)) \cdot \vec u(t).
\end{equation}

Equation \eqnref{PPV} is the phase macromodel of the oscillator; it can
directly capture the phase response under perturbation.
The time-varying periodic vector $\vec v$ inside equation \eqnref{PPV} is known
as the perturbation projection vector (PPV).
It can be numerically calculated from both shooting and harmonic balance.

When the external perturbation $\vec u(t)$ is periodic itself with frequency
$f_1 = 1/T_1$, equation \eqnref{PPV} can be rewritten as
\begin{equation}\eqnlabel{PPVphi}
\frac{d}{dt} \Delta \phi(t) = f_0-f_1 + f_0 \cdot \vec v_{(1)}^T(f_1\cdot t +
\Delta \phi(t)) \cdot \vec u_{(1)}(f_1\cdot t),
\end{equation}
where $\Delta \phi(t) = (f_0-f_1)\cdot t + f_0\cdot \alpha(t)$ ---
when injection locking occurs, it is the phase difference between the
oscillator's response and the periodic perturbation;
$\vec v_{(1)}$ and $\vec u_{(1)}$ are 1-periodic functions ---
$\vec v_{(1)}(t) = \vec v(T_0\cdot t)$,
$\vec u_{(1)}(t) = \vec u(T_1\cdot t)$.

Then we expand 1-periodic functions $\vec v_{(1)}$ and $\vec u_{(1)}$ with
their Fourier coefficients $\{\vec v_{k(1)}\}$ and $\{\vec u_{l(1)}\}$:
\begin{equation}
\vec v_{(1)}(f_1 t + \Delta \phi(t)) = \sum_{k=-\infty}^{k=\infty}
\vec v_{k(1)} \cdot e^{j 2\pi \left(k f_1 t + k \Delta \phi(t) \right)},
\end{equation}

\begin{equation}
\vec u_{(1)}(f_1 t) = \sum_{l=-\infty}^{l=\infty}
\vec u_{l(1)} \cdot e^{j 2\pi \left(l f_1 t \right)}.
\end{equation}

Therefore, \eqnref{PPVphi} can be written as a double summation in the Fourier
domain:
\begin{equation}\eqnlabel{PPVfourier}
\frac{d}{dt} \Delta \phi(t) = f_0-f_1 + f_0
\sum_{k=-\infty}^{k=\infty} \sum_{l=-\infty}^{l=\infty}
\vec v_{k(1)}^T \cdot \vec u_{l(1)} \cdot
e^{j 2\pi \left(k f_1\cdot t + k \Delta \phi(t) + l f_1 t \right)}.
\end{equation}

Simplification of \eqnref{PPVfourier} results in the Generalized Adler's
Equation (GAE) \cite{BhRoASPDAC2009}, and it provides good approximations to
the injection-locked solutions of \eqnref{PPVphi}.

When $\vec u(t)$ contains only the second-order perturbation, \ie, it is
sinusoidal with a frequency of $2f_1$, $\vec u_{2(1)} = \vec u_{-2(1)}^* \neq
0$ are the only non-zero coefficients in $\{\vec u_{l(1)}\}$.
From the analyses in \cite{NeRoDATE2012SHIL}, the larger the magnitude of
$\vec v_{2(1)}$, the more detuning between $f_1$ and $f_0$ the oscillator can
endure while still having an injection-locked solution.

\begin{figure}[htbp]
	\vspace{-0.5em}
	\begin{minipage}{0.49\linewidth}
      \epsfig{file=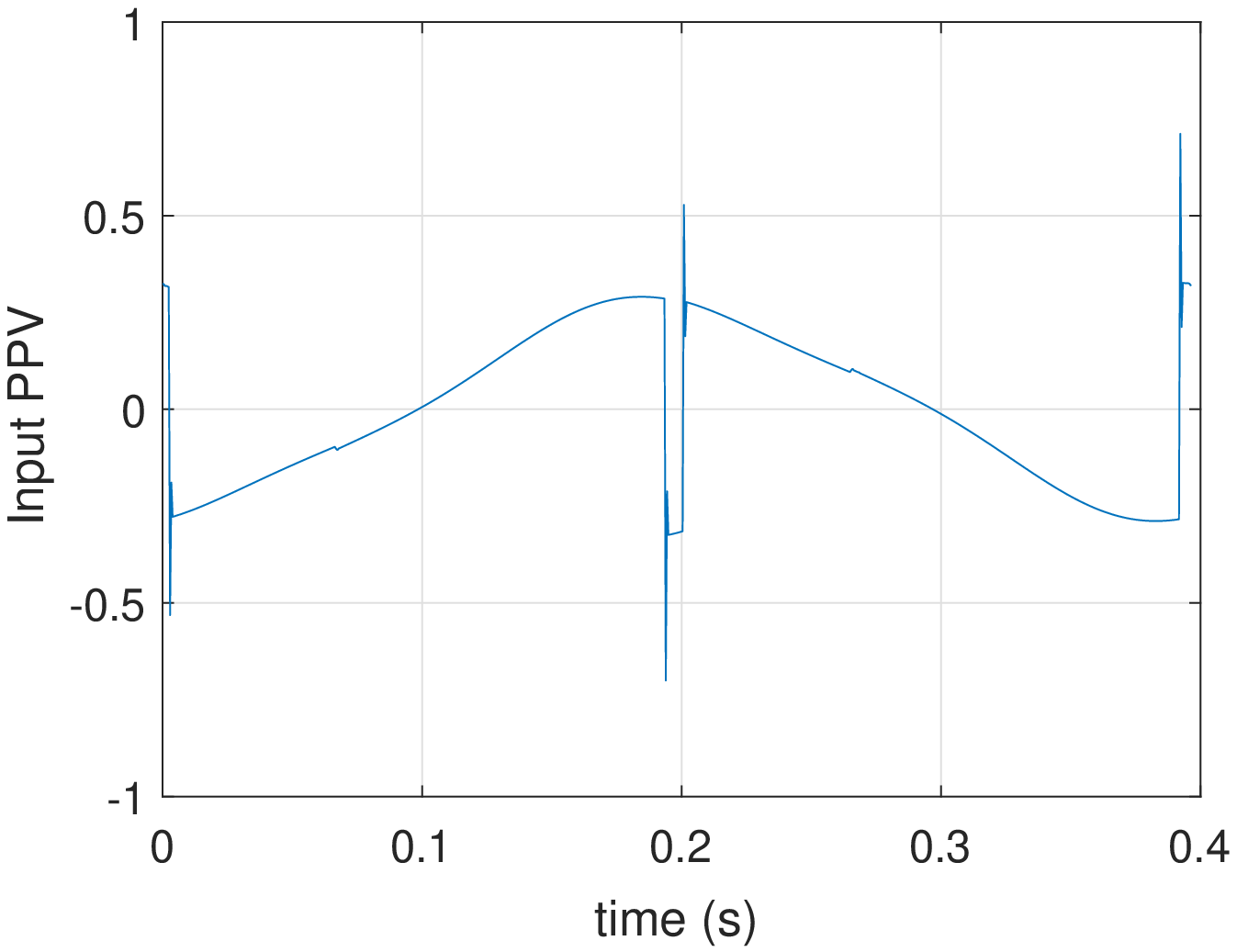,width=\linewidth}
	  \caption{Input PPV of metronome.}\figlabel{PPV-sym-time}
	\end{minipage}
    \hfill
	\begin{minipage}{0.49\linewidth}
      \epsfig{file=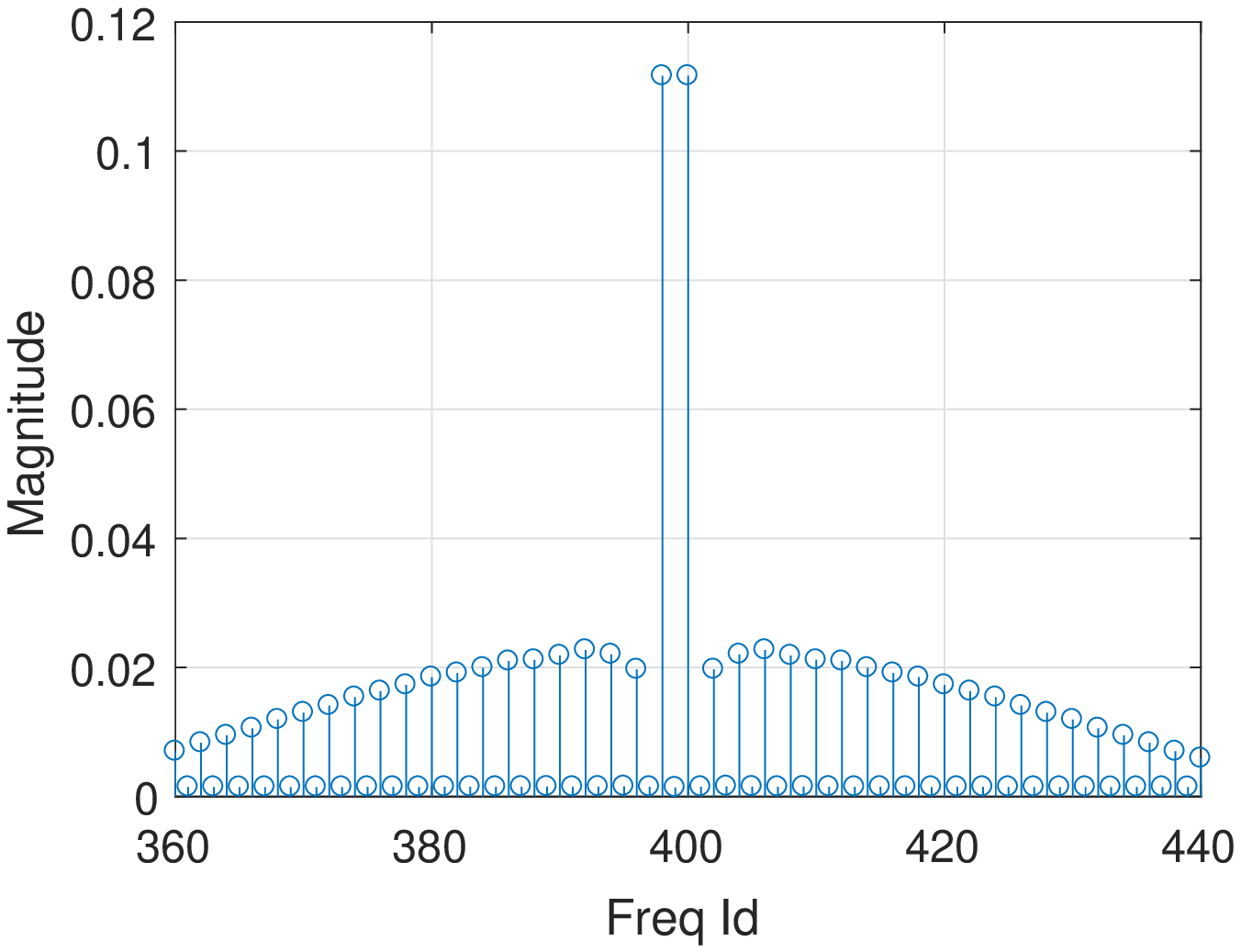,width=\linewidth}
	  \caption{Frequency domain coefficients of metronome's PPV.}\figlabel{PPV-sym}
	\end{minipage}
	\vspace{-0.5em}
\end{figure}

\figref{PPV-sym-time} and \figref{PPV-sym} show the time- and frequency-domain
PPV of a metronome. The metronome model has only one input variable $a$, which
is the horizontal external input acceleration; the PPV becomes a time-varying
scaler that describes the phase's sensitivity to this input.
From \figref{PPV-sym} we see that the PPV has a large coefficient at the
fundamental frequency, indicating that a metronome is prone to regular
first-order IL.
But the second harmonic of the PPV is around $10^{-12}$, in the order of
numerical noise.
In fact, all even harmonics are practically zero, because both the waveforms
and PPV of a metronome are designed to be odd symmetric.
It would be easier to SHIL the metronome with a periodic input $u(t)$ at
$3f_1$, but if we insist on using $2f_1$, the metronome needs to be modified.

\begin{figure}[htbp]
\centering{
	\epsfig{file=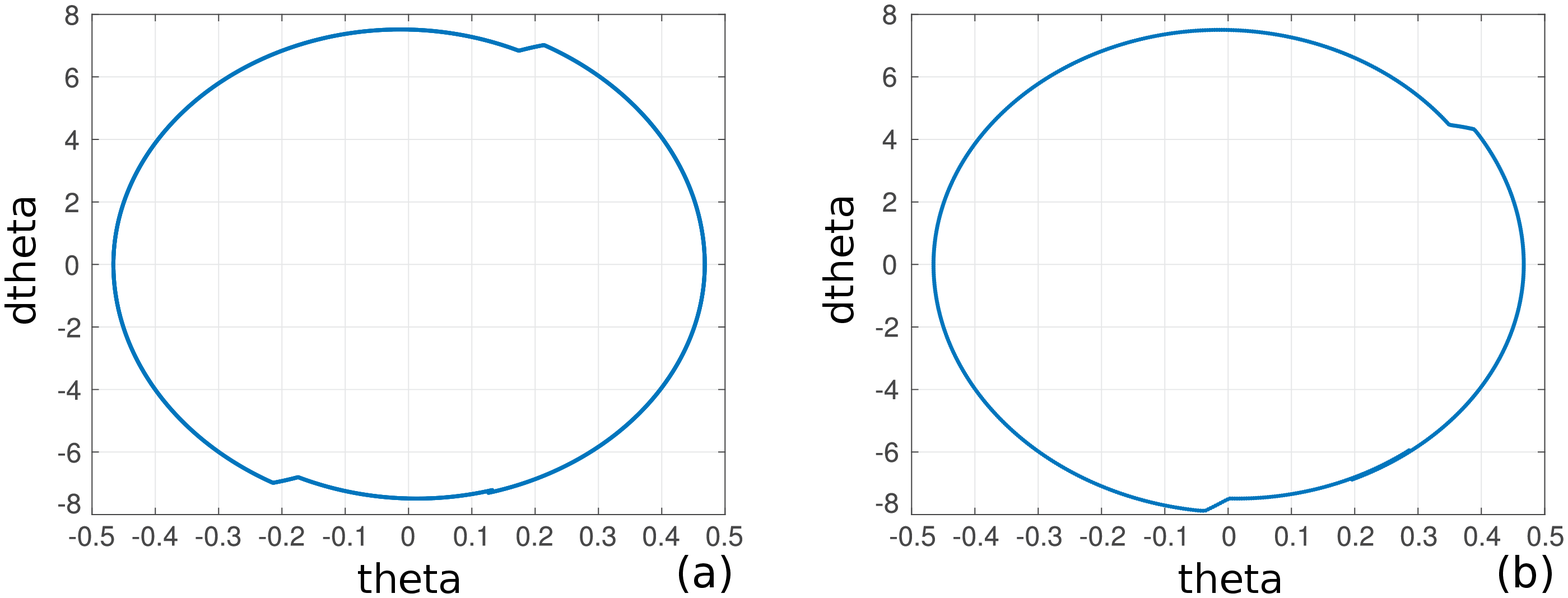,width=0.8\linewidth}
}
\caption{Phase portrait of a metronome's oscillation before and after
    modification.
    \figlabel{metronome-phase-plot}}
\end{figure}

The phase portrait in \figref{metronome-phase-plot} (a) illustrates the
oscillation of a metronome; it is another way of visualizing the $\theta$ and
$\dot \theta$ waveforms in \figref{metronome-waveforms}.
The two small kinks in the loop result from the escapement mechanism, which
accelerates $\dot \theta$ for a short time.
They are symmetric about the origin.
If we open up the metronome, use pliers to rotate the circular plate, we can
adjust the values of $\theta_R$ and $\theta_L$ in \eqnref{gfunc}, thus making
the phase portrait asymmetric, as illustrated in \figref{metronome-phase-plot}
(b).
After this modification, the two ticks generated in each cycle are not equally
separated anymore.
But interestingly, the duration of the metronome oscillation does not change;
the modification does not seem to affect the metronome's energy consumption.

\begin{figure}[htbp]
	\begin{minipage}{0.49\linewidth}
      \epsfig{file=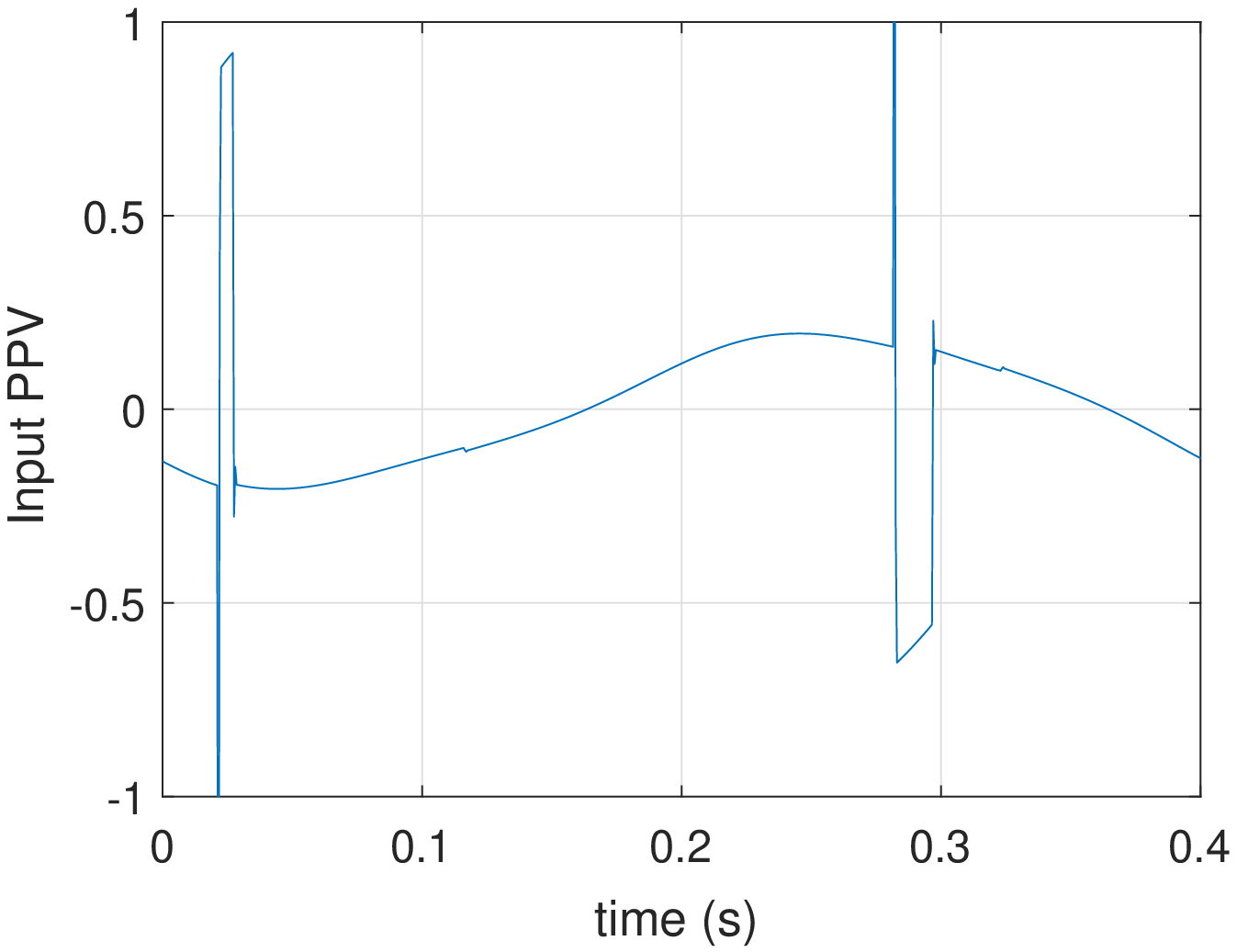,width=\linewidth}
	  \caption{Input PPV of modified metronome.}\figlabel{PPV-asym-time}
	\end{minipage}
    \hfill
	\begin{minipage}{0.49\linewidth}
      \epsfig{file=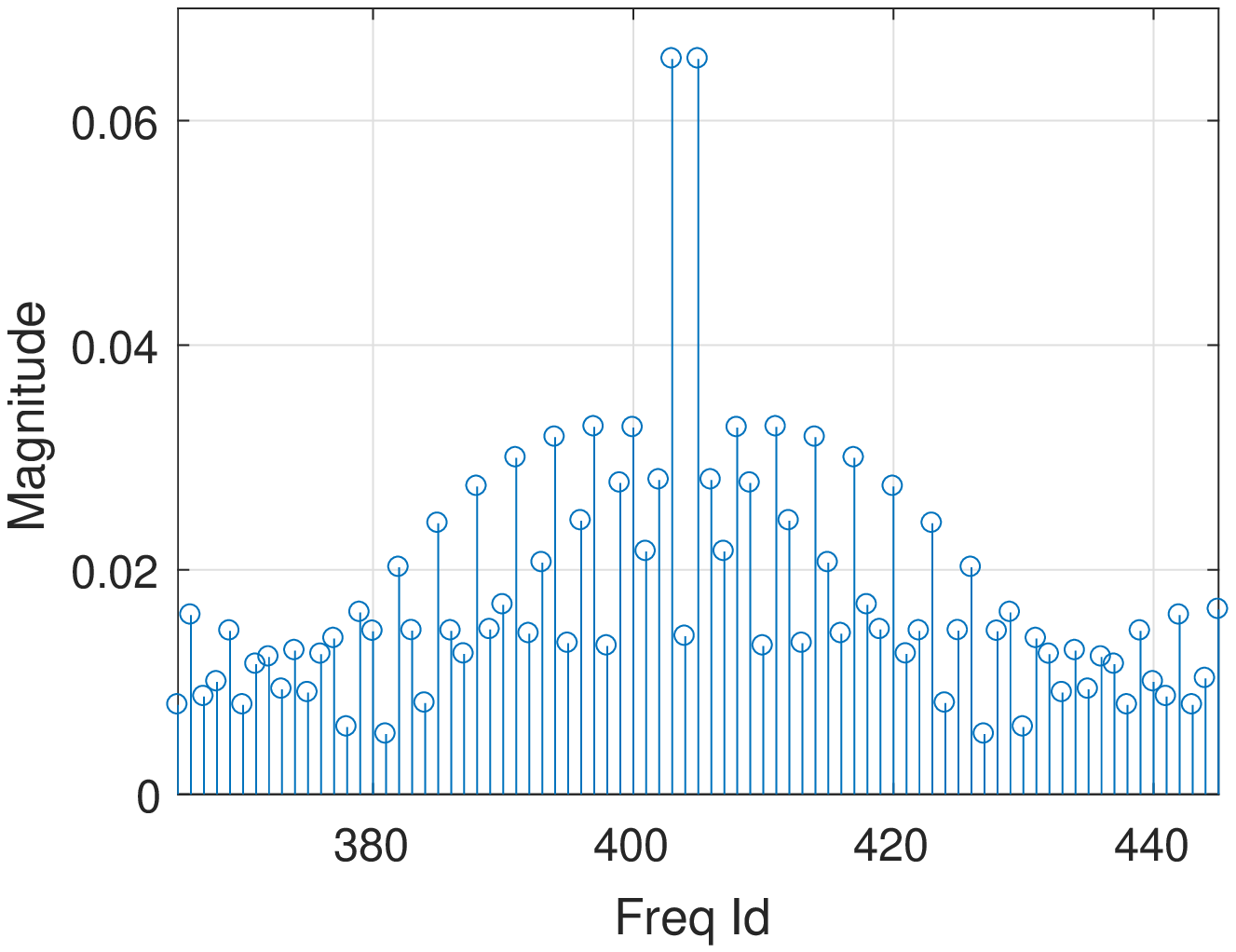,width=\linewidth}
	  \caption{Frequency domain coefficients of modified metronome's PPV.}\figlabel{PPV-asym}
	\end{minipage}
\end{figure}

\figref{PPV-asym-time} and \figref{PPV-asym} show the PPV of the metronome
after modification.
Compared against \figref{PPV-sym-time} and \figref{PPV-sym}, the second
harmonic of the PPV increases from zero to $0.028$, which is now about $43\%$
of the coefficient at fundamental frequency.
The observation of SHIL should now becomes considerably easier; it should be
almost as reproducible as regular IL in metronomes.

\section{\normalfont {\Large Experimental Results}}\seclabel{results}

As described in \secref{intro}, we placed two metronomes on a rolling board ---
one tuned to around 1Hz, the other around 2Hz.
The 1Hz metronome was modified according to \secref{PPV}.
A red sticker was taped to the rod of the 1Hz metronome; a blue one to the 2Hz
metronome.
With a tripod, we recorded their oscillation with a 30Hz frame rate.
The video was imported frame by frame into \MATLAB; an simple algorithm was
used to extract the locations of red and blue stickers from the video.
For every frame, we determined the centers of the two stickers by selecting
fixed numbers of red-most and blue-most pixels and averaging their locations
respectively.
\figref{metronomes-setups-results} (d) shows the oscillation in the two
stickers' x coordinates within a time frame of about 20 seconds.

\begin{figure}[htbp]
\centering{
	\epsfig{file=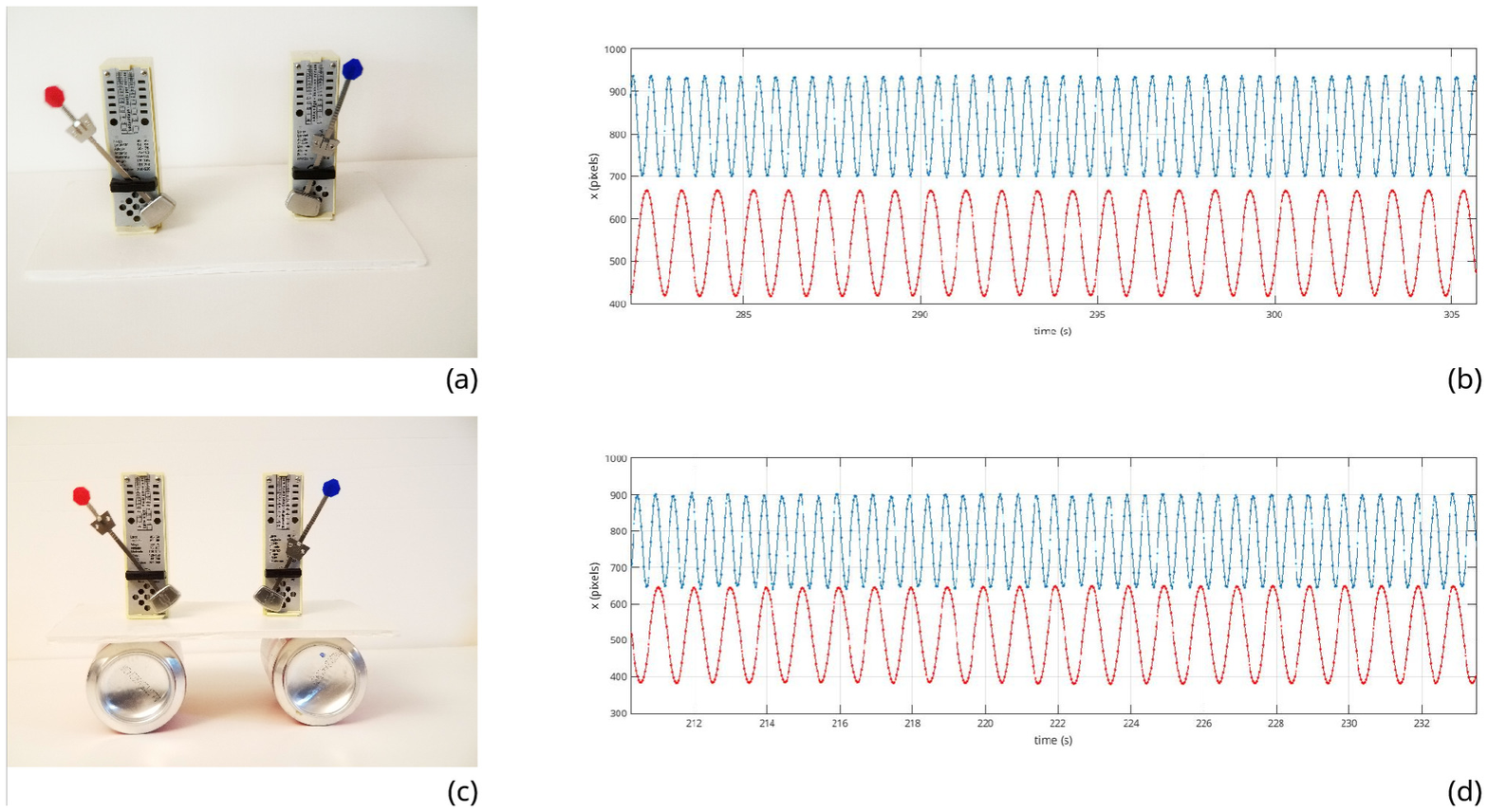,width=1.0\linewidth}
}
\caption{(a) Two metronomes placed on a stationary board; (b) corresponding
waveforms of the tips of their pendulum rods; (c) two metronomes on a rolling board;
(d) their corresponding waveforms.  \figlabel{metronomes-setups-results}}
\end{figure}

Similarly, without retuning the metronomes, we placed them on a stationary
board instead, processed the video and plotted the oscillation in
\figref{metronomes-setups-results} (b).
Throughout the plot, the peaks of the red waveform are almost aligned to every
other valleys of the blue waveform.
However, at the beginning of the plot, the red peak is on the right of the blue
valley, but towards the end of the plot, it has clearly drifted to the left of
the corresponding blue valley.
This is an indication that the two metronomes are not synchronized, whereas in
\figref{metronomes-setups-results} (d), the two waveforms are aligned within
the same duration, indicating the occurrence of SHIL.

\begin{figure}[htbp]
\centering{
	\epsfig{file=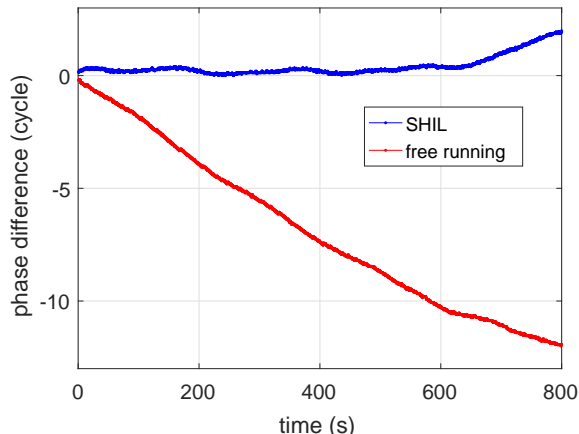,width=0.5\linewidth}
}
\caption{Phase difference between the 2Hz and 1Hz metronomes. The difference is
measured by the number of cycles of the 1Hz metronome. Positive values mean
that the 1Hz metronome is leading the 2Hz one. \figlabel{phase_diff}}
\end{figure}

We can also identify all the peaks in \figref{metronomes-setups-results} (b)
and (d) with the help of a \MATLAB command \texttt{findpeaks()}.
Based on the locations of the peaks, we can plot the phase difference between
the two metronomes in \figref{phase_diff}.
When the two metronomes are free running on a stationary board, the phase
difference keeps increasing.
Specifically, the 1Hz metronome lags the 2Hz one by approximately 10 cycles
after 600 seconds.
But when they are on the rolling board, the phase difference stays constant
around zero for the same duration.
Put in other words, every 2 cycles of the 2Hz metronome align almost perfectly
with 1 cycle of the 1Hz one.
The measurements are not perfectly flat mainly because the spring box does not
generate a constant force, and a metronome's frequency changes slightly as the
spring unrolls.
Also, as the two metronomes oscillate beyond 10 minutes towards the end of
their oscillation, their frequencies drift more and more, and SHIL is
eventually broken.

Furthermore, from the locations of the peaks, we can calculate the frequencies
of the two metronomes.
From the stationary board to the rolling board, the 2Hz metronome maintains the
same frequency at $2f_1 = 1.978$, but the 1Hz one changes its frequency by
about 1\%, from $f_0 = 0.998$ to $f_1 = 0.989$.
This indicates that it is the slower metronome that moves its frequency from
$f_0$ to $f_1$, same as our expectation in \secref{intro}.

\section{\normalfont {\Large Summary}} \seclabel{conclusion}

This paper demonstrates SHIL in mechanical metronomes.
Along the way, we introduce and open-source release compact models for
metronomes that are accurate, physical and can work well in almost all
open-source and commercial simulators.
We also analyze metronomes with PPV-based techniques, which in return guide us
in designing a modification to metronomes to make them more suitable for SHIL.
The phase-locked states from SHIL enable an oscillator to store phase-based
logic values and make almost all oscillators potential candidates for Boolean
computation.
Demonstrating bit storage in a metronome requires another metronome injection
locked to the same frequency to provide a phase reference.
For example, in the setup in this paper, another 1Hz metronomes also locked to
the 2Hz one is needed to distinguish between the two locked states.
This will be part of our future work.

\let\em=\it
\bibliographystyle{unsrt}
{\scriptsize\bibliography{stringdefs,tianshi,jr,PHLOGON-jr}}

\begin{thebibliography}{10}

\bibitem{kinget2002injection}
P.~Kinget, R.~Melville, D.~Long, and V.~Gopinathan.
\newblock {An injection-locking scheme for precision quadrature generation}.
\newblock {\em Solid-State Circuits, IEEE Journal of}, 37(7):845--851, 2002.

\bibitem{Abramovitch2002ACC}
D.~Abramovitch.
\newblock {Phase-locked loops: a control centric tutorial}.
\newblock In {\em Proc. American Control Conference}, volume~1, pages 1 -- 15
  vol.1, May 2002.

\bibitem{tiebout2004cmos}
M.~Tiebout.
\newblock {A CMOS direct injection-locked oscillator topology as high-frequency
  low-power frequency divider}.
\newblock {\em Solid-State Circuits, IEEE Journal of}, 39(7):1170--1174, 2004.

\bibitem{GoTaWeBl1983Lasers}
L.~Goldberg, H.F. Taylor, J.F. Weller, and D.M. Bloom.
\newblock Microwave signal generation with injection-locked laser diodes.
\newblock {\em Electronics Letters}, 19(13):491--493, June 1983.

\bibitem{WaRoUCNC2014PHLOGON}
{T. Wang and J. Roychowdhury}.
\newblock {PHLOGON: PHase-based LOGic using Oscillatory Nanosystems}.
\newblock In {\em Proc. UCNC}, LNCS sublibrary: Theoretical computer science
  and general issues. Springer, July 2014.
\newblock \putDOI{\href{http://dx.doi.org/10.1007/978-3-319-08123-6_29}{DOI
  link.}}

\bibitem{RoPHLOGONarXiVv12014}
J.~Roychowdhury.
\newblock {Boolean Computation Using Self-Sustaining Nonlinear Oscillators}.
\newblock {\em arXiv:1410.5016v1 [cs.ET]}, Oct 2014.
\newblock \putDOI{\href{http://arxiv.org/abs/1410.5016v1/}{arXiv:1410.5016v1.}}

\bibitem{WaRoDAC2015MAPPforPHLOGON}
{T. Wang and J. Roychowdhury}.
\newblock {Design Tools for Oscillator-based Computing Systems}.
\newblock In {\em Proc.\/ IEEE DAC}, pages 188:1--188:6, 2015.
\newblock \putDOI{\href{http://dx.doi.org/10.1145/2744769.2744818}{DOI link.}}

\bibitem{pantaleone2002metronome}
{J. Pantaleone}.
\newblock {Synchronization of Metronomes}.
\newblock {\em American Journal of Physics}, 70(10):992--1000, 2002.

\bibitem{jia2015triplet}
{J. Jia, Z. Song, W. Liu, J. Kurths, J. Xiao}.
\newblock {Experimental Study of the Triplet Synchronization of Coupled
  Nonidentical Mechanical Metronomes}.
\newblock {\em Scientific reports}, 5, 2015.

\bibitem{metronomewebsite}
{Metronome Model Release}.
\newblock Web site:
  \href{https://github.com/TianshiWang/metronome}{https://github.com/TianshiWang/metronome}.

\bibitem{wang2016well}
{T. Wang and J. Roychowdhury}.
\newblock {Well-Posed Models of Memristive Devices}.
\newblock {\em arXiv preprint arXiv:1605.04897}, 2016.

\bibitem{makkar2005friction}
{C. Makkar, W.E. Dixon, W.G. Sawyer, and G. Hu}.
\newblock {A new continuously differentiable friction model for control systems
  design}.
\newblock In {\em Advanced Intelligent Mechatronics. Proceedings, 2005
  IEEE/ASME International Conference on}, pages 600--605. IEEE, 2005.

\bibitem{BhRoASPDAC2009}
P.~Bhansali and J.~Roychowdhury.
\newblock {Gen-Adler: The generalized Adler's equation for injection locking
  analysis in oscillators}.
\newblock In {\em Proc. IEEE ASP-DAC}, pages 522--227, January 2009.
\newblock \href{http://dx.doi.org/10.1109/ASPDAC.2009.4796533}{DOI link.}

\bibitem{NeRoDATE2012SHIL}
{A. Neogy and J. Roychowdhury}.
\newblock {Analysis and Design of Sub-harmonically Injection Locked
  Oscillators}.
\newblock In {\em Proc. IEEE DATE}, Mar 2012.
\newblock \putDOI{\href{http://dx.doi.org/10.1109/DATE.2012.6176677}{DOI
  link.}}

\end{thebibliography}

\appendices
\section{\normalfont Model Code for a Single Metronome}
\applabel{metronome}

 \subsection{\normalfont \texttt{metronomeDAE.m}: model file for metronome DAE in MAPP}
 \applabel{metronome_DAE_code}
 \matlabscript{code/single_metronome_w_input2.m}{lst:metronome_DAE}{\texttt{metronomeDAE.m}}

 \subsection{\normalfont \texttt{metronome.va}: Verilog-A model for a single
 metronome with acceleration input.}
 \applabel{metronome_va_code}
 \verilogascript{code/single_metronome_w_input2.va}{lst:metronome_va}{\texttt{metronome.va}}

\ignore{
 \subsection{\normalfont \texttt{test\_hys.m}: circuit and test script for hys in MAPP}
 \matlabscript{code/test_hys.m}{lst:test_hys_m}{\texttt{test\_hys.m}}

 \subsection{\normalfont \texttt{test\_hys.cir}: circuit and test script for hys in Xyce}
 \verilogascript{code/test_hys.cir}{lst:test_hys_cir}{\texttt{test\_hys.cir}}

 \subsection{\normalfont \texttt{test\_hys.scs}: circuit and test script for hys in \Spectre}
 \verilogascript{code/test_hys.scs}{lst:test_hys_scs}{\texttt{test\_hys.scs}}

 \subsection{\normalfont \texttt{test\_hys.sp}: circuit and test script for hys in \HSPICE}
 \verilogascript{code/test_hys.sp}{lst:test_hys_sp}{\texttt{test\_hys.sp}}

\section{\normalfont Model and Circuit Code for RRAM version 0}
 \subsection{\normalfont \texttt{RRAM\_v0\_ModSpec.m}: model file for RRAM version 0 in MAPP}
 \applabel{RRAM_v0_ModSpec_code}
 \matlabscript{code/RRAM_v0_ModSpec.m}{lst:RRAM_v0_ModSpec}{\texttt{RRAM\_v0\_ModSpec.m}}

 \subsection{\normalfont \texttt{RRAM\_v0.va}: Verilog-A model for RRAM version 0}
 \applabel{RRAM_v0_va_code}
 \verilogascript{code/RRAM_v0.va}{lst:RRAM_v0_va}{\texttt{RRAM\_v0.va}}

 \subsection{\normalfont \texttt{test\_RRAM\_v0.m}: circuit and test script for RRAM version 0 in MAPP}
 \matlabscript{code/test_RRAM_v0.m}{lst:test_RRAM_v0_m}{\texttt{test\_RRAM\_v0.m}}

 \subsection{\normalfont \texttt{test\_RRAM\_v0.cir}: circuit and test script for RRAM version 0 in Xyce}
 \verilogascript{code/test_RRAM_v0.cir}{lst:test_RRAM_v0_cir}{\texttt{test\_RRAM\_v0.cir}}

 \subsection{\normalfont \texttt{test\_RRAM\_v0.scs}: circuit and test script for RRAM version 0 in \Spectre}
 \verilogascript{code/test_RRAM_v0.scs}{lst:test_RRAM_v0_scs}{\texttt{test\_RRAM\_v0.scs}}

 \subsection{\normalfont \texttt{test\_RRAM\_v0.sp}: circuit and test script for RRAM version 0 in \HSPICE}
 \verilogascript{code/test_RRAM_v0.sp}{lst:test_RRAM_v0_sp}{\texttt{test\_RRAM\_v0.sp}}

\section{\normalfont Model Code for Memristor}
 \subsection{\normalfont \texttt{Memristor.m}: model file for memristor ModSpec model in MAPP}
 \applabel{memristor_ModSpec_code}
 \matlabscript{code/Memristor.m}{lst:Memristor}{\texttt{Memristor.m}}

 \subsection{\normalfont \texttt{Memristor.va}: Verilog-A model for Memristor}
 \applabel{memristor_va_code}
 \verilogascript{code/Memristor.va}{lst:Memristor_va}{\texttt{Memristor.va}}

 \subsection{\normalfont \texttt{smoothfunctions.va}: Verilog-A file for smoothing function definitions}
 \applabel{smoothfunctions_va_code}
 \verilogascript{code/smoothfunctions.va}{lst:smoothfunctions}{\texttt{smoothfunctions.va}}
}

\end{document}